\documentclass[reprint,showpacs,showkeys,superscriptaddress,aps,pra]{revtex4-1}

\usepackage{graphicx}
\usepackage{epstopdf}
\usepackage{bm}
\usepackage{amssymb}
\usepackage{amsmath}
\usepackage{bbold}
\usepackage[colorlinks=true, linkcolor=blue]{hyperref}
\usepackage{ulem}
\usepackage{color, soul}
\usepackage{xcolor}
\usepackage{afterpage}
\usepackage{footmisc}
\usepackage{natbib}

\begin{document}

\title{Large spin Hall angle and spin mixing conductance in highly resistive antiferromagnetic Mn\textsubscript{2}Au }
\author{Braj Bhusan Singh}
\email{brajbhusan@niser.ac.in}
\author{Subhankar Bedanta}
\email{sbedanta@niser.ac.in}

\address{Laboratory for Nanomagnetism and Magnetic Materials (LNMM), School of Physical Sciences, National Institute of Science Education and Research (NISER), HBNI, Jatni-752050, India}

\begin{abstract}

Antiferromagnetic materials (AFM) recently have shown interest in the research in spintronics due to its zero stray magnetic field, high anisotropy, and spin orbit coupling. In this context, the bi-metallic AFM Mn\textsubscript{2}Au has drawn attention because it exhibits unique properties and it’s Neel temperature is very high ($T_N$ = 1500 K). Here, we report spin pumping and inverse spin Hall effect (ISHE) investigations in Mn\textsubscript{2}Au/CoFeB bilayer system using ferromagnetic resonance. We found large spin Hall angle $\theta_{SH}$ = 0.22 with comparable spin Hall conductivity i.e. $\sigma_{SH}$ = 1.46 $\times$ $10^5$ ($\hbar/2e$) $\Omega^{-1}m^{-1}$ to the Pt. Further, we have evaluated the effective spin mixing conductance $g_{eff}^{\uparrow \downarrow}$ = 3.27 $\times$ $10^{18}$ $m^{-2}$  and intrinsic spin mixing conductance $g_{r}^{\uparrow \downarrow}$= 8.83$\times$ $10^{18}$ $m^{-2}$  which are higher than the previously reported value (1.40 $\times$ $10^{18}$ $m^{-2}$ for Mn\textsubscript{2}Au/Y\textsubscript{3}Fe\textsubscript{5}O\textsubscript{12}).

\end{abstract}

\maketitle
\section{Introduction }

Antiferromagnetic materials (AFM) have shown significant potential for the applications in spintronics devices [1, 2]. High magnetic anisotropy, zero net magnetization, and hence zero stray magnetic field make them insensitive to the external magnetic field. It helps in the reduction of the bit size in the data storage by removing any cross talk between the stored information unlike ferromagnetic (FM) materials [3]. In addition, they have shown high spin orbit coupling (SOC) which is very useful for the charge to pure spin current conversion or vice versa by spin orbital torque and spin pumping phenomena [4, 5]. Pure spin current is very useful for the fabrication of power efficient spintronics devices [6]. Spin current  can be converted into charge current in a FM/nonmagnetic (NM) heterostructure through a phenomenon known as inverse spin Hall effect (ISHE) [7]. The NM is desired to exhibit high SOC and usually considered to be the heavy metals such as Pt, Ta, W etc. The converted charge current in NM layer can give measurable electric field ($\Vec{E}_{ISHE}$) which is given by [7]:

\begin{equation}\label{q1}
    \Vec{E}_{ISHE} \propto \theta_{SH} \Vec{J}_s \times \Vec{\sigma}
\end{equation}

where 

\begin{equation}\label{q2}
    \Vec{J}_s = \frac{\hbar}{4 \pi}g_{r}^{\uparrow \downarrow}\hat{m} \times \frac{d \hat{m}}{dt}
\end{equation}

where $\theta_{SH}$ is the spin Hall angle which defines spin current ($\Vec{J}_s$) to charge current ($\Vec{J}_c$) conversion efficiency, $\hat{m}$ is the unit vector of magnetization and $\vec{\sigma}$ is the spin matrices governed by the spin polarization direction. Therefore, in order to obtain high $E_{ISHE}$ and hence spin mixing conductance ($g_{r}^{\uparrow \downarrow}$) in a FM/NM heterostructure, $\theta_{SH}$ of the NM needs to be large. The value of $\theta_{SH}$ typically depends on SOC and conductivity of the NM [6, 8]. Ferromagnetic resonance (FMR) based spin pumping experiment is an efficient method to generate spin current by the dynamic transfer of spin angular momentum into NM layer [6, 9–11]. ISHE has been extensively studied in the FM/NM systems where NM are Pt, Ta, and W [12–18]. In recent years there has been extensive search for new materials which are not heavy metals however could exhibit high SOC so that such charge to spin conversion can be efficiently achieved. In this regard topological insulators have been shown to be good candidates which exhibit high SOC due to their metallic surface states [19, 20]. Further AFM materials have been chosen for such charge to spin conversion based spintronics because they also exhibit high SOC. In recent years AFM materials e.g. PtMn, IrMn, FeMn, and PdMn have shown the high ISHE and large value of $\theta_{SH}$ (0.02 – 0.08) in AFM/FM bilayers [21–24]. This encourages to explore other AFM materials for obtaining high $\theta_{SH}$. In this context, Mn\textsubscript{2}Au is a bi-metallic and collinear AFM having very high Neel temperature ($T_N$ = 1500 K) [25–27]. It has been shown that Mn\textsubscript{2}Au exhibits interesting properties like Neel spin orbit torques driven in-plane antiferromagnetic resonance mode [25]. Bodnar \textit{et al}. have shown that in-plane switching of Neel vector in Mn\textsubscript{2}Au by current pulses [28, 29]. X. Chen \textit{et al.} have demonstrated the electric field induced strain switching of Neel spin orbital torque in Mn\textsubscript{2}Au [30]. Most of the reports have focused on switching of Neel spin orbit torque, whereas spin pumping phenomena using Mn\textsubscript{2}Au is less explored. Recently, M. Arana \textit{et al. }have investigated the spin pumping in insulating Mn\textsubscript{2}Au/Y\textsubscript{3}Fe\textsubscript{5}O\textsubscript{12} structure [31]. They observed the value of $\theta_{SH}$= 0.04 and $g_{eff}^{\uparrow \downarrow}$ = 1.40 $\times$ $10^{18}$ $m^{-2}$ . These values are comparable to the Pt [32]. They have prepared thin films of Mn\textsubscript{2}Au by post thermal treatment of Au/Mn/Au layers. It has been found that deposition conditions may change the structural phase and resistivity of the NM material, which are key parameters to observe large value of $\theta_{SH}$. Further high value of $\theta_{SH}$ was observed in case of Pt [32], Ta [33], and W [34] when these NM films were prepared in high resistive beta phase. Considering these aspects,  we have prepared high resistive ( $\sim$ 8 times higher than reported in [31]) Mn\textsubscript{2}Au thin films from a stoichiometric target by sputtering. It should be noted that Y\textsubscript{3}Fe\textsubscript{5}O\textsubscript{12} is an insulating ferrimagnetic and having low magnetic moment, which limits its use for the device applications. Further growth of Y\textsubscript{3}Fe\textsubscript{5}O\textsubscript{12} requires Gd\textsubscript{3}Ga\textsubscript{5}O\textsubscript{12} (GGG) substrate for good lattice matching. However GGG substrates are very expensive. Therefore, it is desired to explore new combinations of Mn\textsubscript{2}Au and other low damping FM materials. In this context CoFeB (CFB) is a low damping material, which usually grows amorphous at room temperature. CFB has been widely used in magnetic tunnel junction based devices [35], which makes it suitable candidate for spin pumping studies. Here, we report the study the spin pumping and ISHE in Mn\textsubscript{2}Au/Co\textsubscript{40}Fe\textsubscript{40}B\textsubscript{20} bilayer. We have observed large value of $\theta_{SH}$ = 0.22 which is about five times higher than reported in reference 31 [31]. We have also evaluated spin interface transparency and spin Hall conductivity in Mn\textsubscript{2}Au/Co\textsubscript{40}Fe\textsubscript{40}B\textsubscript{20}, which are new information for a Mn\textsubscript{2}Au based system.

\section{EXPERIMENTAL DETAILS}

We fabricated Mn\textsubscript{2}Au (10 nm)/Co\textsubscript{40}Fe\textsubscript{40}B\textsubscript{20} (5 nm)/TaO\textsubscript{x}(2 nm) bilayer thin film by dc magnetron sputtering. The base pressure of the vacuum system (manufactured by Mantis Deposition Ltd., UK) was better than 5 $\times$ $10^{-8}$ mbar. During growth of Mn\textsubscript{2}Au thin film the substrate temperature was kept at $200^{\circ}$ C. Subsequently, the sample has been cooled down to room temperature in the vacuum and then Co\textsubscript{40}Fe\textsubscript{40}B\textsubscript{20} thin film was grown. A thin Ta layer was also deposited on top of the Co\textsubscript{40}Fe\textsubscript{40}B\textsubscript{20} layer to protect it from oxidation. In our case, Ta layer naturally oxidize to form $TaO_{x}$ layer. We have also deposited a control sample having structure Co\textsubscript{40}Fe\textsubscript{40}B\textsubscript{20} (5 nm)/ TaO\textsubscript{x}(2 nm) for reference in same growth conditions. Further, in order to confirm the crystalline quality and phase of Mn\textsubscript{2}Au, we deposited a 20 nm thick Mn\textsubscript{2}Au thin film. The deposition conditions were kept same which were used for the fabrication of Mn\textsubscript{2}Au (10 nm)/Co\textsubscript{40}Fe\textsubscript{40}B\textsubscript{20} (5 nm)/TaO\textsubscript{x}(2 nm) sample. All samples were deposited on MgO(100) substrate. For obtaining smooth surfaces and removing water vapors, the MgO(100) substrate was heat treated at 600 $\circ$ C/1hr before deposition of Mn\textsubscript{2}Au thin film. We performed x-ray diffraction ($\lambda$ = 1.5418 $\AA$) in $\theta$-2$\theta$ geometry (in a Rigaku diffractometer) for characterizing crystalline quality of the samples. X-ray reflectivity (XRR) was measured for thickness and interface roughness measurements. FMR measurements have been performed in the frequency range 5-17 GHz to evaluate the damping properties, where sample was kept in the flip chip manner on a 200 $\mu$m wide coplanar wave guide (CPW). ISHE measurements have been performed by connecting a nanovoltmeter (Keithley 2182A) at the two ends of the sample by silver paste at a fixed frequency. The schematic of the measurements configuration of the FMR and ISHE are shown in Fig. \ref{fig1}(c). The detailed methodology of FMR and ISHE have been described in our earlier report [20, 45].

\section{Results}

\subsection*{Crystalline quality}
\begin{figure*}[ht]
	\centering
	\includegraphics[height=145mm]{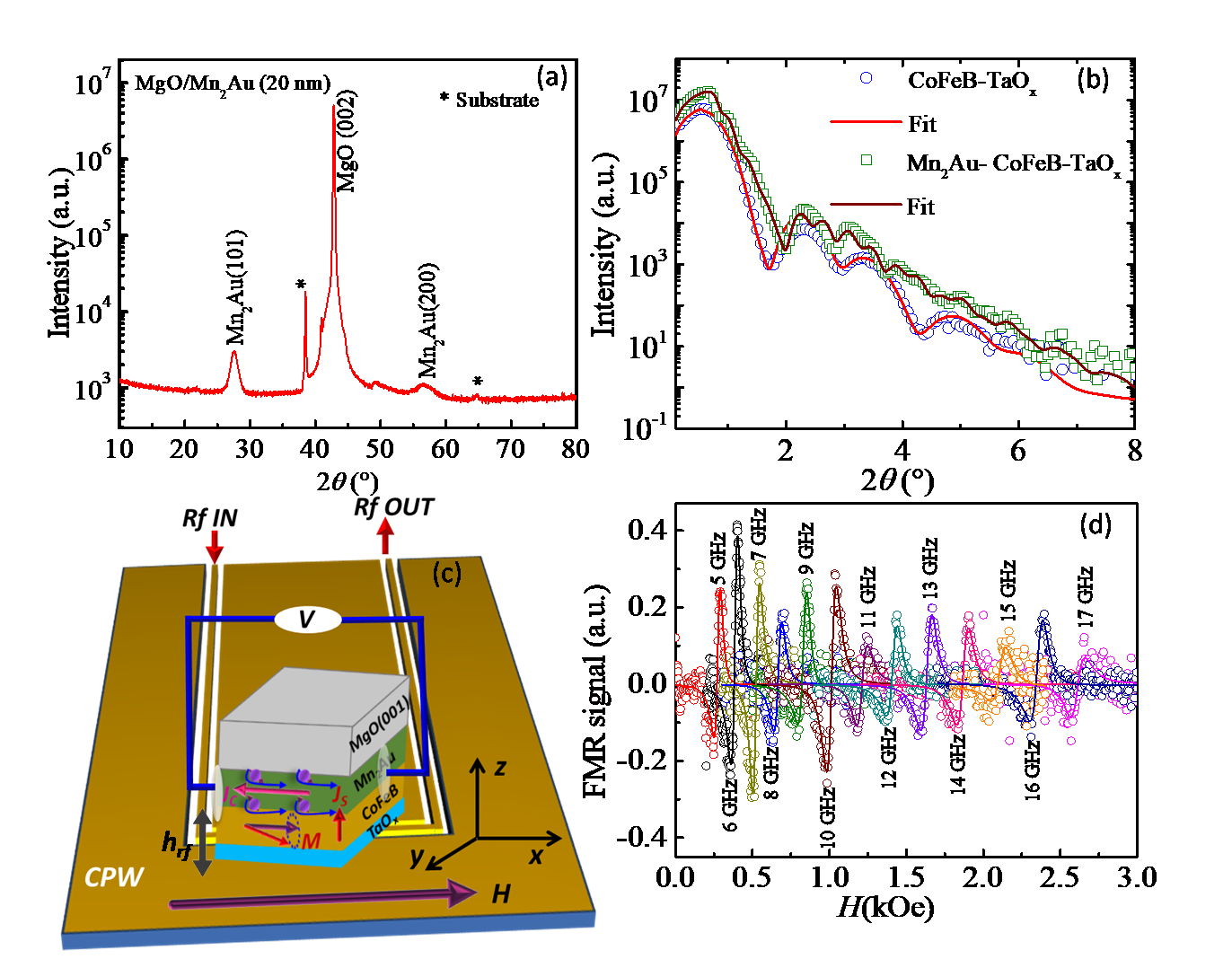}
	\caption{Structural properties and ferromagnetic resonance measurements (a) X-ray diffraction data measured in $\theta$-2$\theta$ geometry for the sample MgO (100) /Mn\textsubscript{2}Au (20 nm) (b) XRR measurements data for the samples Mn\textsubscript{2}Au (10 nm)/Co\textsubscript{40}Fe\textsubscript{40}B\textsubscript{20} (5 nm)/TaO\textsubscript{x}(2 nm) (square symbol) and Co\textsubscript{40}Fe\textsubscript{40}B\textsubscript{20} (5 nm)/TaO\textsubscript{x}(2 nm) (circle symbol). Solid lines are best fit to the experimental data. (c) Experimental set-up for FMR and ISHE measurements. The \textit{rf} field ($h_{rf}$) which is perpendicular to the applied magnetic field (H) defines the direction of spin current ($J_s$) (d) FMR spectra of the sample Mn\textsubscript{2}Au (10 nm)/Co\textsubscript{40}Fe\textsubscript{40}B\textsubscript{20} (5 nm)/TaO\textsubscript{x}(2 nm) measured in the frequency range of 5-17 GHz. Open symbols are experimental data, while solid lines are the best fit using equation (\ref{q3}).}
	\label{fig1}
\end{figure*}

Fig. \ref{fig1}(a) shows the x-ray diffraction of the single layer Mn\textsubscript{2}Au film of 20 nm thickness deposited on MgO (100) substrate. We have observed the diffraction peaks of Mn\textsubscript{2}Au corresponding to (101) and (200) planes. We did not observe any diffraction peak in the grazing incidence x-ray diffraction. It indicates the texture growth of Mn\textsubscript{2}Au thin films. Fig. \ref{fig1}(b) shows the XRR measurements data (open symbols) for the samples Mn\textsubscript{2}Au (10 nm)/Co\textsubscript{40}Fe\textsubscript{40}B\textsubscript{20} (5 nm)/TaO\textsubscript{x}(2 nm) and Co\textsubscript{40}Fe\textsubscript{40}B\textsubscript{20} (5 nm)/TaO\textsubscript{x}(2 nm). Oscillations corresponding to Mn\textsubscript{2}Au and Co\textsubscript{40}Fe\textsubscript{40}B\textsubscript{20} (5 nm) are clearly observed. The solid lines are best to the experimental data using X’pert reflectivity software. The interface roughness of Mn\textsubscript{2}Au/CoFeB layer is found to be 0.39 $\pm$ 0.01 nm, which indicates the smother interfaces. 
 
\subsection*{Spin pumping and inverse spin Hall effect measurement}
\begin{figure}[h!]
	\centering
	\includegraphics[width=0.45\textwidth]{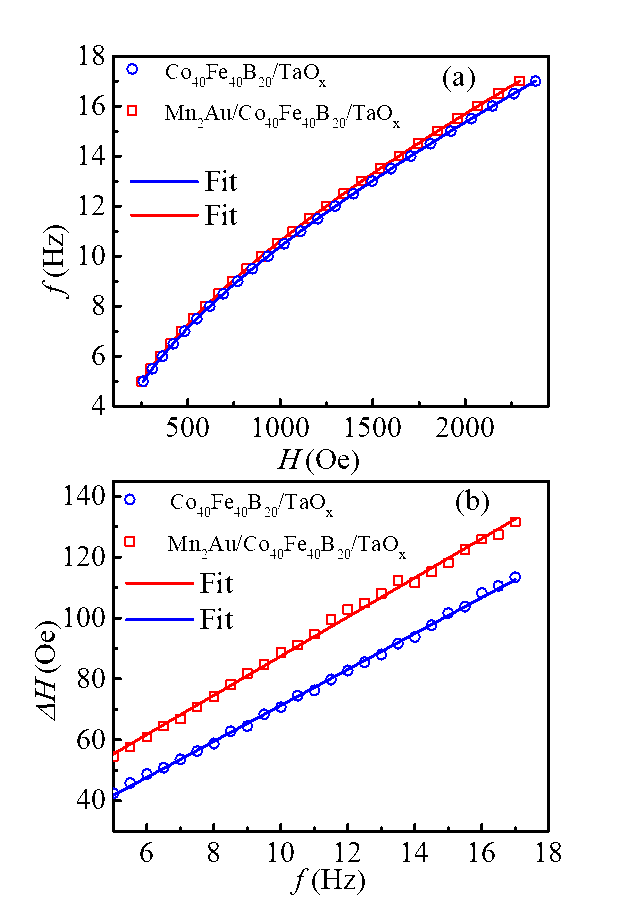}
	\caption{(a) Frequency (\textit{f}) versus $H_{res}$ (b) \textit{$\Delta$H} versus \textit{f} for the sample Mn\textsubscript{2}Au (10 nm)/Co\textsubscript{40}Fe\textsubscript{40}B\textsubscript{20} (5 nm)/TaO\textsubscript{x}(2 nm) (square symbol) and Co\textsubscript{40}Fe\textsubscript{40}B\textsubscript{20} (5 nm)/TaO\textsubscript{x}(2 nm) (circle symbols). Solid lines are best fit to the equation (\ref{q4}) and (\ref{q6}).}
	\label{fig2}
\end{figure}

We performed measurements of FMR spectra in the frequency range of 5-17 GHz as shown in Fig. \ref{fig1}(d) (Open symbols). In order to evaluate the values of resonance field (\textit{$H_{res}$} ) and line width (\textit{$\Delta$H} ) from the FMR spectra, we fitted experimental data by Lorentzian function:
\begin{equation}\label{q3}
\begin{aligned}
\textit{FMR Signal} = K_1\frac{4\Delta{H}(H-H_{res})}{[4(H-H_{res})^2+\Delta{H}^2]^2}\\-K_2\frac{\Delta{H}^2-4(H-H_{res})^2}{[4(H-H_{res})^2+\Delta{H}^2]^2}+ offset
\end{aligned}
\end{equation}

where $K_1$ and $K_2$ are coefficient of the anti-symmetric and the symmetric components, respectively.
Figs. \ref{fig2}(a) and (b) show the plots of frequency (\textit{f}) versus \textit{$H_{res}$} and\textit{ $\Delta{H}$} versus \textit{f}, respectively. In order to find the values of effective demagnetization (4$\pi$$M_{eff}$) and gyromagnetic ratio ($\gamma$); experimental data as shown in Fig. \ref{fig2}(a) was fitted to Kittel’s equation [36] which is given by:

\begin{equation}\label{q4}
    \it{f}=\frac{\gamma}{2 \pi} \sqrt{(H_K+H_{res})(H_K +H_{res} + 4 \pi M_{eff})}
\end{equation}

where 
\begin{equation}\label{q5}
    4 \pi M_{eff} = 4 \pi M_s + \frac{2K_S}{M_st_{FM}}
\end{equation}

and  $t_{FM}$, $H_K$, $K_s$, $M_s$ are thickness of FM layer,anisotropy field, perpendicular surface magnetic anisotropy constant, and saturation magnetization, respectively. The values of the damping constant ($\alpha$) was calculated by fitting experimental data of Fig. \ref{fig2}(b) using the following equation [37]:

\begin{equation}\label{q6}
    \Delta H = \Delta H_0 +\frac{4 \pi \alpha \it{f}}{\gamma}
\end{equation}

where \textbf{$\Delta H_0$} is the inhomogeneous line width broadening. It depends on the magnetic homogeneity of the sample. This equation gives effective value of damping constant. It may included other effects e.g. interface effect, impurity, magnetic proximity effects (MPE) etc., which also can enhance the value of $\alpha$ of the system. Therefore, we can write the total $\alpha$ as: 
\begin{equation}\label{q7}
    \alpha = \alpha_{int} + \alpha_{impurity} + \alpha_{mpe} + \alpha_{sp}
\end{equation}

where $\alpha_{int}$ is the intrinsic damping, and $\alpha_{impurity}$, $\alpha_{mpe}$, and $\alpha_{sp}$ are the contribution from impurity, magnetic proximity effect (MPE), and spin pumping to the damping constant, respectively [38].

The linear behaviour of \textit{$\Delta H$} vs $\textit{f}$ plots implies the better homogeneity in our samples and it rules out any possibility of two magnon scattering in our samples. 

\begin{figure}[h!]
	\centering
	\includegraphics[width=0.45\textwidth]{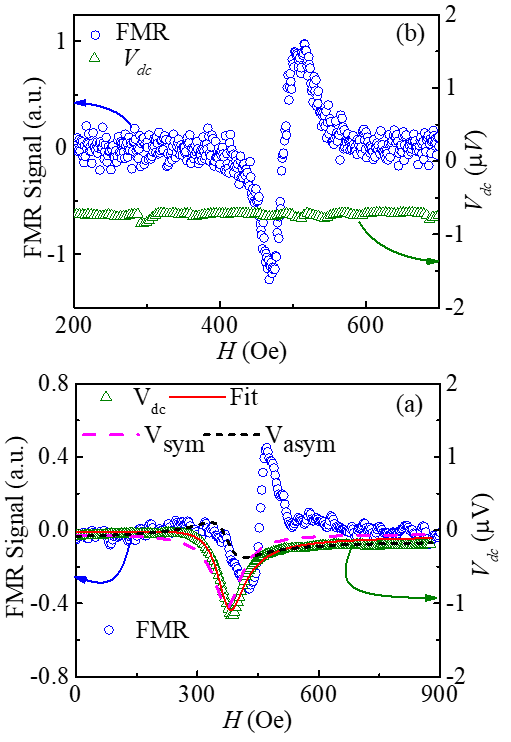}
	\caption{$V_{dc}$ (circle symbol) and FMR signal (square symbol) versus applied magnetic field (\textit{H}) for the samples (a) Mn\textsubscript{2}Au (10 nm) (10 nm)/Co\textsubscript{40}Fe\textsubscript{40}B\textsubscript{20} (5 nm)/TaO\textsubscript{x}(2 nm) (b) (a) Co\textsubscript{40}Fe\textsubscript{40}B\textsubscript{20} (5 nm)/TaO\textsubscript{x}(2 nm). Solid line is the best fit to the equation (\ref{q8}), while dotted and dashed lines correspond to $V_{asym}$ and $V_{sym}$ contributions, respectively. }
	\label{fig3}
\end{figure}

The evaluated value of $\alpha$ of Mn\textsubscript{2}Au (10 nm)/Co\textsubscript{40}Fe\textsubscript{40}B\textsubscript{20} (5 nm)/TaO\textsubscript{x}(2 nm) sample is found to be 0.00991$\pm$ 0.00001. This value of $\alpha$ is higher than the value of Co\textsubscript{40}Fe\textsubscript{40}B\textsubscript{20} (5 nm)/TaO\textsubscript{x}(2 nm) (0.00896$\pm$ 0.00001), which indicates that there may be the presence of spin pumping. However, we cannot rule out the enhancement of $\alpha$ due to other effects arising from interfaces. In order to confirm the spin pumping, we performed measurement of ISHE by connecting a nanovoltmeter across the sample, as shown in Fig. \ref{fig1}(c). The measured dc voltage ($V_{dc}$) along with FMR signal plotted with magnetic field are shown in Fig. \ref{fig3}. In order to distinguish symmetric ($V_{sym}$) and antisymmetric components ($V_{asym}$) experimental data of $V_{dc}$ (circle symbol) was fitted with Lorentzian function (solid line) which is [39]:

\begin{equation}\label{q8}
\begin{aligned}
    V_{dc} = V_{sym} \frac{(\Delta H)^2}{(H-H_{res})^2+(\Delta H)^2}+ \\
    V_{asym} \frac{2 \Delta H (H - H_{res})}{(H-H_{res})^2+(\Delta H)^2}
    \end{aligned}
\end{equation}

\begin{figure}[h!]
	\centering
	\includegraphics[width=0.45\textwidth]{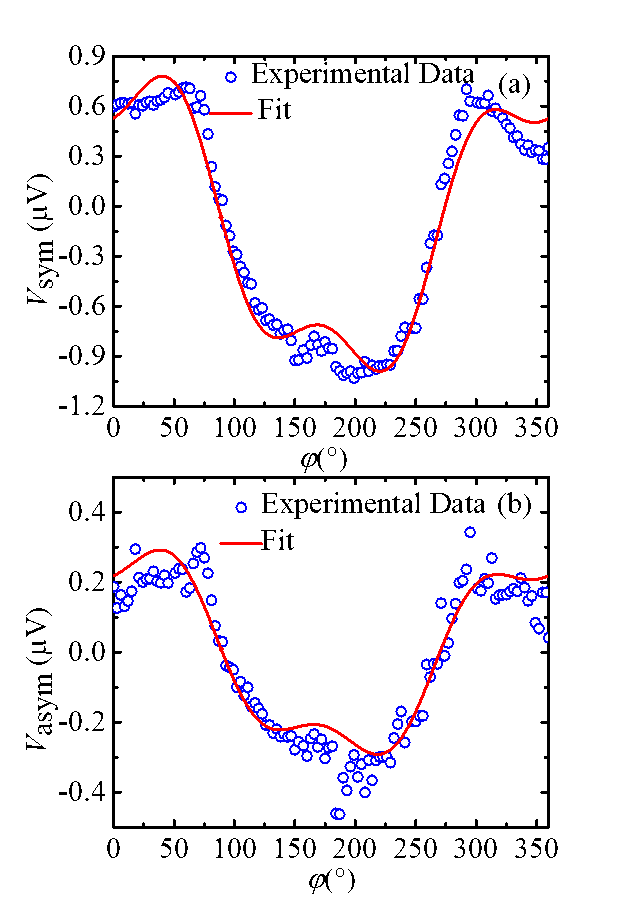}
	\caption{Angle dependent $V_{sym}$ (a) and $V_{asym}$ (b) components for the sample Mn\textsubscript{2}Au (10 nm)/Co\textsubscript{40}Fe\textsubscript{40}B\textsubscript{20} (5 nm)/TaO\textsubscript{x}(2 nm).}
	\label{fig4}
\end{figure}

Solid lines are the fits to the experimental data. The $V_{sym}$ consists of major contribution from spin pumping, while minor contributions from anomalous Hall effect (AHE), and anisotropic magnetoresistance (AMR) effects. It should be noted that the AHE contribution is zero here if the \textit{rf} field and H are perpendicular to each other, which is the case in our measurement. Whereas the AHE and AMR are the major contributions in the $V_{asym}$ components. Fig. \ref{fig4} also shows the plot of $V_{sym}$ (dashed line) and $V_{asym}$ (dotted line) separately for the sample Mn\textsubscript{2}Au (10 nm)/Co\textsubscript{40}Fe\textsubscript{40}B\textsubscript{20} (5 nm)/TaO\textsubscript{x}(2 nm).
In order to quantify spin pumping and other spin rectification contributions in-plane angle dependent measurements of $V_{dc}$ were performed at the interval of $3^{\circ}$ (Fig. \ref{fig4}). It is a well-established method to decouple the individual components from the measured voltage [38–40]. The model given by Harder \textit{et al.} [41] has considered the rectification effects i.e., parallel AMR ($V_{asym/sym}^{AMR ||}$) and perpendicular AMR ($V_{asym/sym}^{AMR \perp}$) to the applied \textit{rf} field and the AHE contribution due to the FM layer. The relation between the measured voltage and those rectification effects are as follows [39]:
\begin{equation}\label{q9}
\begin{aligned}
 V_{asym}= V_{AHE}cos(\phi) sin (\it{\Phi})+ 
    V_{asym}^{AMR \perp} cos 2(\phi)sin(\it{\Phi})\\+ 
   V_{asym}^{AMR ||}sin2(\phi)cos(\phi)
   \end{aligned}
 \end{equation}

\begin{equation}\label{q10}
\begin{aligned}
    V_{sym}= V_{sp}cos^3(\phi)+V_{AHE}cos(\phi)cos(\it{\Phi)}
    +\\ V_{sym}^{AMR \perp} cos 2(\phi)cos(\phi)
    + V_{sym}^{AMR ||}sin2(\phi)cos(\phi)
    \end{aligned}
\end{equation}

$V_{AHE}$ and $V_{sp}$ correspond to the AHE voltage and the spin pumping contributions, respectively. $\phi$ is the angle between applied \textit{H} and the rf magnetic field which is $90^{\circ}$ in the present measurement. Further the AMR contribution also can be quantified by the following formula(39):

\begin{equation}\label{q11}
  V_{AMR}=\sqrt{(V_{asym}^{AMR \perp,||})^{2}+(V_{sym}^{AMR \perp,||})^{2}}  
\end{equation}

The \textit{$V_{asym}^{AMR \perp,||}$} and \textit{$V_{sym}^{AMR \perp,||}$} are evaluated from the in-plane angle dependent $V_{dc}$ measurements by fitting those values by equations (\ref{q9}) and (\ref{q10}), respectively. The values of $V_{sp}$, $V_{AMR}^{\perp}$  and $V_{AMR}^{||}$ were found to be 2.21 $\pm$ 0.05, 1.57 $\pm$ 0.05, and 0.14 $\pm$ 0.02 $\mu$V, respectively. Higher value of $V_{sp}$ in comparison to other effects, indicates that the spin pumping is the dominating mechanism for in the enhancement of $\alpha$ in Mn\textsubscript{2}Au (10 nm)/Co\textsubscript{40}Fe\textsubscript{40}B\textsubscript{20} (5 nm)/TaO\textsubscript{x}(2 nm).

\begin{figure}[h!]
	\centering
	\includegraphics[width=0.45\textwidth]{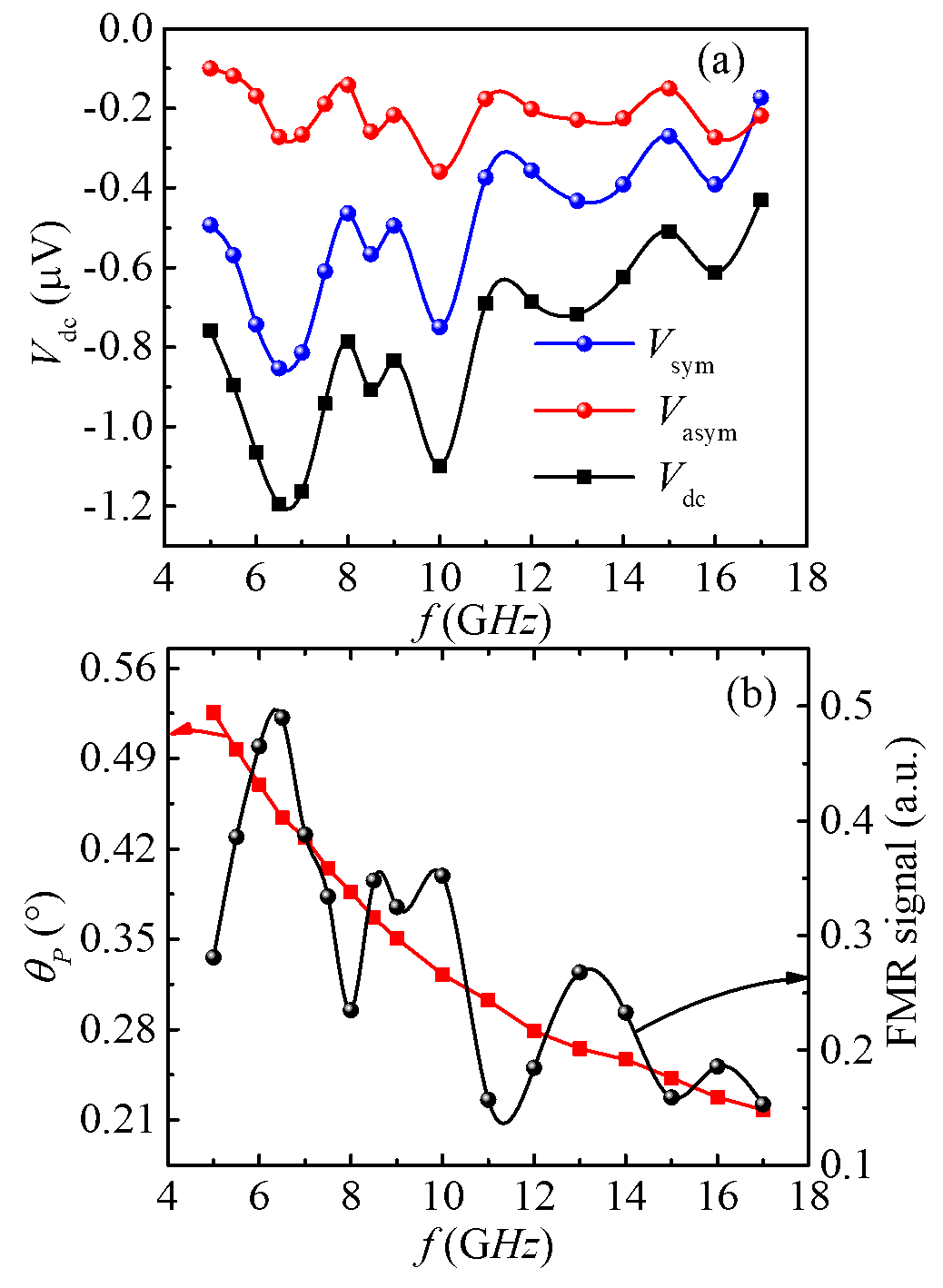}
	\caption{Frequency dependence of spin pumping voltage (a) Frequency dependent $V_{dc}$  ,$V_{sym}$ and $V_{asym}$ components. (b) Frequency dependent of precession angle ($\theta_P$) and FMR signal strength for sample Mn\textsubscript{2}Au (10 nm)/Co\textsubscript{40}Fe\textsubscript{40}B\textsubscript{20} (5 nm)/TaO\textsubscript{x}(2 nm).}
	\label{fig5}
\end{figure}

\begin{figure}
	\includegraphics[width=0.45\textwidth]{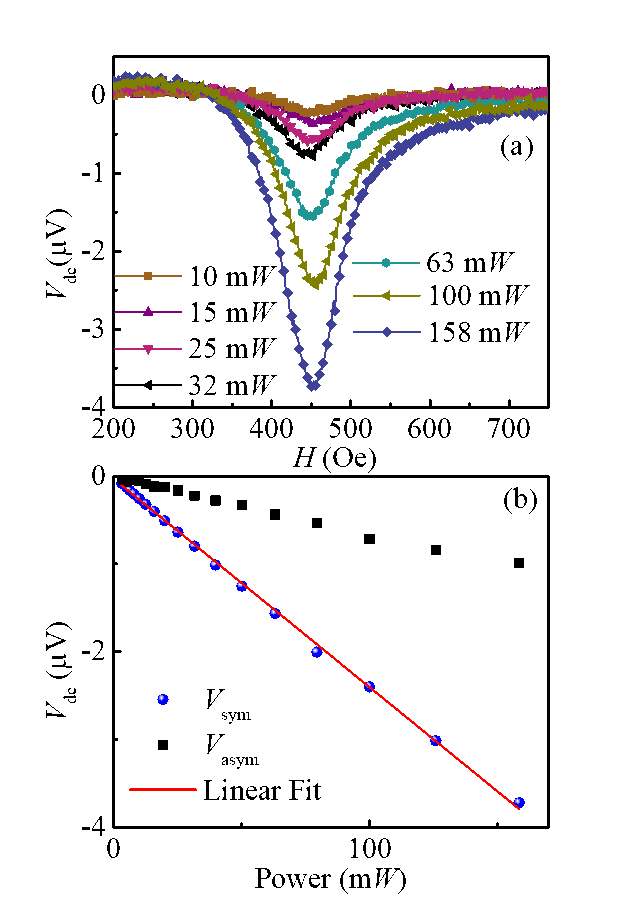}
	\caption{\textbf{Effect of power on spin pumping voltage.} Power dependence of the $V_{dc}$ (a); $V_{sym}$ and $V_{asym}$ (b). Solid line is the linear fit.}
	\label{fig6}
\end{figure}

To understand the effect of precession frequency on the ISHE, we performed frequency dependent measurement in the range 5 to 17 GHz at an interval of 0.5 GHz. Frequency dependent $V_{dc}$  is shown in figure \ref{fig5}(a). It is observed that the measured voltage is highest at 6.5 GHz frequency. The value of $V_{dc}$  at the frequencies 7 and 10 GHz are having similar value, but, it is lower in the case of 8 and 9 GHz. Further, we plotted frequency dependent $V_{sym}$ and $V_{asym}$ which is shown in Fig. \ref{fig6}(b). It is observed that the values of $V_{sym}$ and $V_{asym}$ are higher when the value of $V_{dc}$  is larger. We did not observe any clear trend of decreasing $V_{sym}$ and $V_{asym}$ with increase in frequency as observed by Vlaminck \textit{et al}[42]. It is known that at higher frequencies the losses in CPW increases and also the precession cone angle ($\theta_P$) decreases, therefore, it is expected that at higher frequencies the value of $V_{dc}$  will increase. In order to confirm this, we evaluated the values of $\theta_P$=$h_{rf}$/$\Delta$H , where $\Delta$H is the linewidth of FMR spectra, at different frequencies [42]. Figure \ref{fig5}(b) shows the graph $\theta_P$ versus \textit{f} (square symbols). It is observed that the values of $\theta_P$ are decreasing continuously with frequency, however, measured $V_{dc}$ is not monotonously decreasing (Fig.\ref{fig5}(a)). It is known that the magnitude of $V_{dc}$ should be higher at large cone angle due to second order dependence of $J_S$ $\propto$ sin2$\theta_P$ [42]. It is noted that the values of $\theta_P$ are very small compared to other reports [42, 43], which is less than $1^{\circ}$, which will not affect substantially. the $V_{dc}$. Therefore, it is concluded that the $\theta_P$ is not a major reason for the fluctuation in $V_{dc}$ with frequency. In order to understand this we plotted FMR signal (peak value) versus frequency, which is shown in Fig. \ref{fig5}(b) (circle symbols). We compared the behavior of $V_{dc}$ and FMR signal with frequency (Fig. \ref{fig5}(a, b)) and observed that the behavior is mimicking each other. Therefore, it can be concluded that strength of FMR signal is the primary reason for the variation in $V_{dc}$   and hence $V_{sym}$ and $V_{asym}$ with frequency.

Spin pumping should increase with increase in \textit{rf} power ($P_{rf}$) due to the linear dependence of $h_{rf}$ on $P_{rf}$ ($h_{rf}$ $\propto$ $\sqrt{P_{rf}}$) [42]. Hence, we performed the power dependence measurements of $V_{dc}$ which are shown in the Fig.\ref{fig6}(a). It is showing that the $V_{dc}$ is increasing with power, which again indicates the presence of spin pumping. The evaluated $V_{sym}$ and $V_{asym}$ components are presented in the Fig.\ref{fig6}(b). It is clearly showing that $V_{sym}$ component is rapidly increasing with power in comparison of $V_{asym}$ component. Therefore, it is concluded that the spin pumping is the dominating phenomenon in our Mn\textsubscript{2}Au (10 nm)/Co\textsubscript{40}Fe\textsubscript{40}B\textsubscript{20} (5 nm)/TaO\textsubscript{x}(2 nm) system.

Effective spin mixing conductance (\textit{$g_{\it{eff}}^{\uparrow \downarrow}$}) is crucial factor to understand the spin transport across the FM-NM interfaces. The value of \textit{$g_{\it{eff}}^{\uparrow \downarrow}$} can be calculated by the following expression using damping constant [6]:

\begin{equation}\label{q12}
g_{eff}^{\uparrow \downarrow}=\frac{\Delta\alpha 4\pi M_{s}t_{FM}}{g\mu_{B}} 
\end{equation}

where\textit{$\Delta\alpha$}, \textit{$t_{FM}$}, \textit{$\mu_{B}$}, \textit{g} are the change in the $\alpha$ due to spin pumping, the thickness of CoFeB layer, Bohr magneton, and Lande g- factor (2.1), respectively. The value of $M_S$ was 1069 emu/cc which was evaluated by measuring the sample in a superconducting quantum interference device (SQUID) based magnetometer. We found that the value of \textit{$g_{\it{eff}}^{\uparrow \downarrow}$} is 3.27$\pm$ 0.02 $\times$ $10^{18}$ $m^{-2}$. This value is higher than the reported value (1.40 $\times$ $10^{18}$ $m^{-2}$) in the case of Mn\textsubscript{2}Au/$Y_3Fe_5O_{12}$ system [31]. 

In addition to the \textit{$g_{\it{eff}}^{\uparrow \downarrow}$}, spin interface transparency (T) is another parameter which is useful for spin-orbit torque-based devices. The value of T is affected by the electronic structure matching of FM and NM layers. The following expression is used to calculate T [44]:

\begin{equation}\label{q13}
    T = \frac{g_{r}^{\uparrow \downarrow} tanh(\frac{t_{NM}}{2 \lambda_{NM}})}{g_{r}^{\uparrow \downarrow} coth(\frac{t_{NM}}{\lambda_{NM}})+\frac{h\sigma_{NM}}{2e^2\lambda_{NM}}}
\end{equation}
where $g_{r}^{\uparrow \downarrow}$ is the intrinsic spin mixing conductance which is given by

\begin{equation}\label{q14}
    g_{r}^{\uparrow \downarrow} = g_{eff}^{\uparrow \downarrow}\frac{\frac{\sigma_{NM}h}{\lambda_{NM}2e^2}}{\frac{\sigma_{NM}h}{\lambda_{NM}2e^2}-g_{eff}^{\uparrow \downarrow}}
\end{equation}

 $\sigma_{NM}$ is the conductivity of Mn\textsubscript{2}Au layer. $\lambda_{NM}$ is the spin diffusion length of Mn\textsubscript{2}Au, (here we took 1.6 nm from the literature [31]). The value of $g_{r}^{\uparrow \downarrow}$ was evaluated to be (8.83 $\pm$ 0.02) $\times$ $10^{18}$ $m^{-2}$.We found the value of T to be 0.61 $\pm$ 0.02. This value is comparable to the values people have reported for Pt and others heavy metal systems [44].
Further we have calculated the value of $\theta_{SH}$ for Mn\textsubscript{2}Au (10 nm) by using the following expression [6]:
\begin{equation}\label{q15}
\begin{aligned}
    J_s \approx (\frac{g_{eff}^{\uparrow \downarrow}\hbar}{8\pi})(\frac{\mu_0 h_{rf}\gamma}{\alpha})^2\times\\
    [\frac{\mu_0 M_s\gamma+\sqrt{(\mu_0 M_s\gamma)^2+16(\pi f)^2}}{(\mu_0 M_s\gamma)^2+16(\pi f)^2}](\frac{2e}{\hbar})
    \end{aligned}
\end{equation}

\begin{equation}\label{q16}
    E_{ISHE}=(\frac{w_y}{\sigma_{FM} t_{FM}+\sigma_{NM} t_{NM}}) \theta_{SH} l_{sd}^{NM}tanh(\frac{t_{NM}}{2 \lambda_{NM}}) J_s
\end{equation}
The resistivity of the samples was measured using the four probe technique. The $\rho_{Mn_2Au}$ and $\rho_{CoFeB}$ were obtained to be 1.50$\times$$10^{-6}$ $\Omega-m$ and 2.31$\times$$10^{-6}$ $\Omega-m$ respectively. $\sigma$ corresponds to the conductivity of the individual layers. The rf field ($\mu_0 h_{rf}$) and CPW transmission line width ($w_y$) for our measurement are 0.05 mT (at 15 dBm rf power) and 200 $\mu$m, respectively. 

The estimation of $\theta_{SH}$ is found to be 0.22 $\pm$ 0.01, which is one order higher than the value reported by Arana \textit{et al.} [31] and comparable to the corresponding values for Pt. It is highest value of $\theta_{SH}$ compared to other collinear antiferromagnetic systems. Resistivity of the NM layer play critical role in order to obtain high value $\theta_{SH}$ as evidenced by equation \ref{q16}. It has been shown in the case of Pt [32], Ta [33], and W [34] that highly resistive phase shows the large value of $\theta_{SH}$ due to shorter life time of carrier. Therefore, in our case we have grown highly resistive Mn\textsubscript{2}Au thin films compared to reported by Arana \textit{et al.} [31], which is caused to obtain large value of $\theta_{SH}$. Further, it is noted that that the value of $\theta_{SH}$ critically depends on the spin diffusion length $\lambda_{NM}$, which we took from the literature. Therefore, we have also calculated the $\theta_{SH}$ for different values of $\lambda_{NM}$ and observed that for $\lambda_{NM}$ = 10 nm, it $\theta_{SH}$ $\sim$ 0.076, which is still two times the value reported by Rana \textit{et al.} [31]. We have also evaluated spin Hall conductivity $\sigma_{SH}$ = 1.46 $\times$ $10^5$ ($\hbar$/2e) $\Omega^{-1}m^{-1}$which is comparable to the values for Pt [32].

\section{Conclusions}

We have investigated the inverse spin Hall effect and spin pumping in the Mn\textsubscript{2}Au/CoFeB system. In order to disentangle various spin rectification effects we have performed the angle dependent ISHE. It is observed that in our sample spin pumping is the dominant phenomenon. Further, we have evaluated the effective mixing conductance which is found to be higher than the reported values for Mn\textsubscript{2}Au. Also, we have calculated the spin Hall angle which found to be 0.22 $\pm$ 0.01, which is probably highest value so far in case of any reported collinear antiferromagnetic material. Therefore, Mn\textsubscript{2}Au is a suitable candidate for antiferromagnetic spintronics and work as a replacement for Pt and other heavy metal.

\section*{ACKNOWLEDGEMENT}

The authors acknowledge department of atomic energy (DAE), Govt. of India, for the financial support for the experimental facilities. BBS acknowledges department of science and technology (DST), Govt. of India for INSPIRE faculty fellowship.

\section*{REFERENCES}
\begin{enumerate}
\item V. Baltz, A. Manchon, M. Tsoi, T. Moriyama, T. Ono, Y. Tserkovnyak. \textit{Rev. Mod. Phys}. \textbf{90}, 015005 (2018).
\item O. Gomonay, T. Jungwirth, J. Sinova. \textit{{physica status solidi (RRL)} – Rapid Research Letters}.\textbf{ 11}, 1700022 (2017).
\item 	A. Manchon, J. Železný, I. M. Miron, T. Jungwirth, J. Sinova, A. Thiaville, K. Garello, P. Gambardella. \textit{Rev. Mod. Phys.} \textbf{91}, 035004 (2019).
\item  X. F. Zhou, J. Zhang, F. Li, X. Z. Chen, G. Y. Shi, Y. Z. Tan, Y. D. Gu, M. S. Saleem, H. Q. Wu, F. Pan, C. Song. \textit{Phys. Rev. Applied}. \textbf{9}, 054028 (2018).
\item	J. Železný, Y. Zhang, C. Felser, B. Yan, \textit{Phys. Rev. Lett.} \textbf{119}, 187204 (2017)..
\item	Y. Tserkovnyak, A. Brataas, G. E. W. Bauer, B. I. Halperin, \textit{Rev. Mod. Phys}. \textbf{77}, 1375–1421 (2005).
\item	E. Saitoh, M. Ueda, H. Miyajima, G. Tatara, \textit{Appl. Phys. Lett.} \textbf{88}, 182509 (2006).
\item	Y. Tserkovnyak, A. Brataas, G. E. W. Bauer, \textit{Phys. Rev. Lett}. \textbf{88}, 117601 (2002).
\item	H. Nakayama, K. Ando, K. Harii, T. Yoshino, R. Takahashi, Y. Kajiwara, K. Uchida, Y. Fujikawa, E. Saitoh, \textit{Phys. Rev. B.} \textbf{85}, 144408 (2012).
\item	J.-C. Rojas-Sánchez, N. Reyren, P. Laczkowski, W. Savero, J.-P. Attané, C. Deranlot, M. Jamet, J.-M. George, L. Vila, H. Jaffrès, \textit{Phys. Rev. Lett}. \textbf{112}, 106602 (2014).
\item	S. Keller, J. Greser, M. R. Schweizer, A. Conca, V. Lauer, C. Dubs, B. Hillebrands, E. Th. Papaioannou, \textit{Phys. Rev. B.} \textbf{96}, 024437 (2017).
\item	J. Sinova, S. O. Valenzuela, J. Wunderlich, C. H. Back, T. Jungwirth, \textit{Rev. Mod. Phys.} \textbf{87}, 1213–1260 (2015).
\item	A. Hoffmann, \textit{IEEE Transactions on Magnetics}. \textbf{49}, 5172–5193 (2013). 
\item	S. O. Valenzuela, M. Tinkham, \textit{Nature}. \textbf{442}, 176–179 (2006).
\item	E. Saitoh, M. Ueda, H. Miyajima, G. Tatara, \textit{Appl. Phys. Lett.} \textbf{88}, 182509 (2006).
\item	O. Mosendz, J. E. Pearson, F. Y. Fradin, G. E. W. Bauer, S. D. Bader, A. Hoffmann, \textit{Phys. Rev. Lett.} \textbf{104}, 046601 (2010).
\item	A. Azevedo, L. H. Vilela-Leão, R. L. Rodríguez-Suárez, A. F. Lacerda Santos, S. M. Rezende, \textit{Phys. Rev. B.} \textbf{83}, 144402 (2011).
\item	L. Liu, C.-F. Pai, Y. Li, H. W. Tseng, D. C. Ralph, R. A. Buhrman, \textit{Science}. \textbf{336}, 555–558 (2012)
\item  A. R. Mellnik, J. S. Lee, A. Richardella, J. L. Grab, P. J. Mintun, M. H. Fischer, A. Vaezi, A. Manchon, E.-A. Kim, N. Samarth, D. C. Ralph,\textit{ Nature}. \textbf{511}, 449–451 (2014).
\item	B. B. Singh, S. K. Jena, M. Samanta, K. Biswas, B. Satpati, S. Bedanta, \textit{physica status solidi (RRL) – Rapid Research Letters}. \textbf{13}, 1800492 (2019).
\item	W. Zhang, M. B. Jungfleisch, F. Freimuth, W. Jiang, J. Sklenar, J. E. Pearson, J. B. Ketterson, Y. Mokrousov, A. Hoffmann, \textit{Phys. Rev. B.} \textbf{92}, 144405 (2015).
\item	W. Zhang, M. B. Jungfleisch, W. Jiang, J. E. Pearson, A. Hoffmann, F. Freimuth, Y. Mokrousov, \textit{Phys. Rev. Lett.} \textbf{113}, 196602 (2014).
\item J. B. S. Mendes, R. O. Cunha, O. Alves Santos, P. R. T. Ribeiro, F. L. A. Machado, R. L. Rodríguez-Suárez, A. Azevedo, S. M. Rezende, \textit{Phys. Rev. B.} \textbf{89}, 140406 (2014).
\item J. Zhou, X. Wang, Y. Liu, J. Yu, H. Fu, L. Liu, S. Chen, J. Deng, W. Lin, X. Shu, H. Y. Yoong, T. Hong, M. Matsuda, P. Yang, S. Adams, B. Yan, X. Han, J. Chen, \textit{Science Advances.} \textbf{5}, eaau6696 (2019
\item N. Bhattacharjee, A. A. Sapozhnik, S. Yu. Bodnar, V. Yu. Grigorev, S. Y. Agustsson, J. Cao, D. Dominko, M. Obergfell, O. Gomonay, J. Sinova, M. Kläui, H.J. Elmers, M. Jourdan, J. Demsar, Neel\textit{ Phys. Rev. Lett}. \textbf{120}, 237201 (2018).
\item	S. Khmelevskyi, P. Mohn, \textit{Appl. Phys. Lett.} \textbf{93}, 162503 (2008).
\item 	V. M. T. S. Barthem, C. V. Colin, H. Mayaffre, M.-H. Julien, D. Givord, \textit{Nat Commun}. \textbf{4}, 1–7 (2013).
\item S. Y. Bodnar, L. Šmejkal, I. Turek, T. Jungwirth, O. Gomonay, J. Sinova, A. A. Sapozhnik, H.-J. Elmers, M. Kläui, M. Jourdan, \textit{ Nat Commun.} \textbf{9}, 1–7 (2018).	
\item M. Meinert, D. Graulich, T. Matalla-Wagner, \textit{Phys. Rev. Applied.} \textbf{9}, 064040 (2018).
\item X. Chen, X. Zhou, R. Cheng, C. Song, J. Zhang, Y. Wu, Y. Ba, H. Li, Y. Sun, Y. You, Y. Zhao, F. Pan, \textit{Nat. Mater.} \textbf{18}, 931–935 (2019).	
\item M. Arana, M. Gamino, E. F. Silva, V. M. T. S. Barthem, D. Givord, A. Azevedo, S. M. Rezende, \textit{Phys. Rev. B}. \textbf{98}, 144431 (2018).
\item L. Zhu, L. Zhu, M. Sui, D. C. Ralph, R. A. Buhrman, \textit{Science Advances.} \textbf{5}, eaav8025 (2019).
\item L. Liu, C.-F. Pai, Y. Li, H. W. Tseng, D. C. Ralph, R. A. Buhrman, \textit{Science}. \textbf{336}, 555–558 (2012).
\item 	C.-F. Pai, L. Liu, Y. Li, H. W. Tseng, D. C. Ralph, R. A. Buhrman, \textit{Appl. Phys. Lett.} \textbf{101}, 122404 (2012).
\item 	S. S. P. Parkin, C. Kaiser, A. Panchula, P. M. Rice, B. Hughes, M. Samant, S.-H. Yang, \textit{Nat Mater}. \textbf{3}, 862–867 (2004).
\item C. Kittel, \textit{Phys. Rev.} \textbf{73}, 155–161 (1948).	
\item  B. Heinrich, J. F. Cochran, R. Hasegawa, \textit{Journal of Applied Physics}. \textbf{57}, 3690–3692 (1985).

\item 	A. Conca, S. Keller, L. Mihalceanu, T. Kehagias, G. P. Dimitrakopulos, B. Hillebrands, E. Th. Papaioannou, \textit{Phys. Rev. B.} \textbf{93}, 134405 (2016).

\item A. Conca, B. Heinz, M. R. Schweizer, S. Keller, E. Th. Papaioannou, B. Hillebrands, \textit{Phys. Rev. B.} \textbf{95}, 174426 (2017).

\item M. Harder, Y. Gui, C.-M. Hu, \textit{Physics Reports}. \textbf{661}, 1–59 (2016).

\item M. Harder, Z. X. Cao, Y. S. Gui, X. L. Fan, C.-M. Hu, \textit{Phys. Rev. B.} \textbf{84}, 054423 (2011).

\item 	V. Vlaminck, J. E. Pearson, S. D. Bader, A. Hoffmann, \textit{Phys. Rev. B.} \textbf{88}, 064414 (2013).

\item 	S. Gupta, R. Medwal, D. Kodama, K. Kondou, Y. Otani, Y. Fukuma, \textit{Appl. Phys. Lett}. \textbf{110}, 022404 (2017).

\item W. Zhang, W. Han, X. Jiang, S.-H. Yang, S. S. P. Parkin, \textit{Nature Physics.} \textbf{11}, 496–502 (2015).

\item B. B. Singh, S. K. Jena, S. Bedanta, \textit{J. Phys. D: Appl. Phys}. \textbf{50}, 345001 (2017).

\end{enumerate}

\end{document}